\documentclass[12pt]{article}

\usepackage{array,graphicx}
\usepackage[dvipsnames]{xcolor}

\usepackage[pdftex,%
plainpages=false,%
linktocpage=true,%
a4paper=true,%
colorlinks=true,%
linkcolor=NavyBlue,%
citecolor=BrickRed,%
urlcolor=NavyBlue,%
bookmarks=true]{hyperref}

\title{WordPress on AWS: a Communication Framework}
\author{Michael Soltys and Katharine Soltys}
\date{\today}

\begin{document}
\maketitle

\begin{abstract}
Every organization needs to communicate with its audience, and social
media is an attractive and inexpensive way to maintain dialogic
communication. About 1/3 of the Internet web pages are powered by
WordPress, and about a million companies have moved their IT
infrastructure to the AWS cloud. Together, AWS and WordPress offer an
attractive, effective and inexpensive way for companies, both large
and small, to maintain their presence on the web.
\end{abstract}

\tableofcontents

\section{Introduction}

This paper summarizes a course titled {\em Online Communication and
Society}, taught at California State University Channel Islands with
code COMP 347. This class starts from the following premise: 
\begin{quote}
{\em
you have been hired by a company with a small Communication budget,
but ambitious plans. You have been tasked with setting up an effective
web presence; in this role you have to combine both your CS/IT skills,
as well as your Communication savvy. The decision has been made to
deploy the web page as WordPress hosted on Amazon Web Services
(AWS), integrated with social media, as well as robust Analytics to
measure the effectiveness of your communication campaigns.
}
\end{quote}
This paper, which reflects the content of the course, covers how to
accomplish this mission.  

The material is aimed a junior (3rd year) students in Computer Science
and IT, but it covers all the content from the beginning, and thus it
is accessible to Business and Communication students as well. Of
course, the more Computer Science / IT maturity a student has, the
more they will be able to draw out of this important and topical
material.

\subsection{The Cloud with AWS}

As discussed in \cite{aws-soltys-cloudify}, the Cloud is a new
paradigm of computing. What we mean by ``The Cloud'' is a way of
provisioning IT services that involves accessing them through the
Internet, where they are reachable {\em on-demand} in a {\em
pay-as-you-go} manner.  There are various providers of Cloud services,
where the top three are Amazon Web Services (AWS) with the largest
market share, followed by Microsoft Azure and Google Cloud.

What is attractive about The Cloud is that a business can deploy a
state of the art online presence without making capital investments,
that is purchasing hardware, ``rack-stack-cool-power'' a data center,
and hire an IT department.  The cloud's ``pay-as-you-go'' model scales
according to need, and therefore a business does not have to provision
for peak usage. 

AWS (\cite{aws-soltys-cloudify}) works well with universities as it
has a well developed academic interface, in the form of {\em AWS
Academy} and {\em AWS Educate}, as well as its significant dominance
of the cloud market, which works to the advantage of graduating
students, in search of a job, who have AWS expertise.  About a
million companies use AWS, including giants such as Netflix and
Dropbox, as well as start-ups and non-profits.

This paper describes how to build a scalable, flexible and
fault-tolerant web site using the WordPress application running on the
AWS cloud platform.  In itself, building a web site is not difficult;
but it is another matter to build a web site that {\em scales}, i.e.,
is able to serve a fluctuating number of visitors, {\em flexible}, can
be quickly and easily updated with content and in fact works as a {\em
Content Management System} (CMS), and is {\em fault-tolerant}, i.e.,
when something breaks, and something always breaks, it is still able
to perform its function adequately.

It is possible to use free AWS Educate accounts to learn and practice
all that is presented in this paper. Those who wish to turn that
knowledge into practice can save their {\em Infrastructure as Code}
(IaC), a great advantage of the Cloud, and port it as a \verb|json| or
\verb|yaml| file to another environment.

\subsection{WordPress}

WordPress (WP) is a free and open-source {\em Content Management System}
(CMS) based on PHP and MySQL. It includes a {\em plugin architecture}
and a {\em template system}. While WP is often associated with
blogging, it supports other types of web content including
traditional mailing lists and forums, media galleries, and online
stores.  WP is by far the most used CMS by websites, as can be
seen here \cite{wordpress-cms}. 

As WP is written in PHP and MySQL, familiarity with those two
languages can help with a deeper customization of your site, but it is
not required to understand and implement the material discussed in
this paper. Also, even without knowledge of PHP, but with some
familiarity of programming and scripting, it is possible to make
tweaks to the WP code. To give an example of this, consider
this case: when a WP page is password protected the title shows
as ``Protected: Title of the page.'' The word ``Protected'' can be
confusing to some, and the WP administrator may wish to change
it to something else, e.g., ``Password Protected: Title of the page.''
To accomplish that, simply go to the file
\verb|wp-includes/post-template.php| and replace the first line given
below with the second:
\begin{verbatim}
$prepend = __( 'Protected: %s' );
$prepend = __( 'Password Protected: %s' );
\end{verbatim}
In the author's current version of WP\footnote{Version 5.4},
this is line 127 in \verb|post-template.php|. This example
was put here to illustrate that deeper customization of WP can
be accomplished with little experience with PHP. That said, PHP makes
up over 80\%\ of all server-side code out there, and it is not too
difficult to pick up, and hence a web site administrator is encouraged
to learn it.

In order to host WP, you will need to be running an EC2
instance, which is a virtual machine in the cloud, with a setup that
is called a ``LAMP'' (an acronym for Linux, Apache, MySQL and PHP).
This will be explained in Section~\ref{sec:lamp}. Once you have an EC2
instance with LAMP, you will download WP to it, and proceed
from there as outlined in Section~\ref{sec:wordpress}. Note that
WP is free (the software, that is), and you can download it to
your machine (\verb|wget https://wordpress.org/latest.tar.gz|) and
examine at will before doing an actual installation in the cloud.

Note that it is also possible to host a WP blog on
\url{wordpress.com}; this is a paid service, as it provides a
host machine and ability to register a domain, as well as the
WP installation. It is a managed service, i.e., {\em Software
as a Service} (SaaS), whereas in this paper we examine a {\em Platform
as a Service} (PaaS) installation. Finally, check \cite{aws-wp-blog,
aws-wp-best} for more information on PaaS installations of WP
on AWS, and in particular the \verb|yaml| template for deploying WP on
AWS in \cite{wp-github}.

\section{Working online}

Many professionals work from home these days, a trend accelerated by
the COVID-19 crisis, when most of the nation was working from home
in order to practice ``social isolation.'' It is fundamental for
remote work (and study) to be well versed in the tools that enable it.
Note that these tools are deployed in the Cloud, and hence this course
will in fact have the added benefit of teaching students how to work
productively in an online environment. In this section we are going to
examine the tools used in this online course, which are also the
pre-eminent tools used in the workforce.

Slack and Microsoft Teams are the top work-place messaging systems.
Both applications group workers into ``channels,'' and offer wide
integration with other tools, such as, GitHub, Dropbox and Dropbox
Paper, Google Drive, and many others, as well as teleconferencing.
They support {\em Markdown} which is a lightweight markup language
with a plain-text formatting syntax. In Markdown one can specify
bold-face or italics text with simple tags, such as \verb|*bold*| and
\verb|_italics_|.  See \cite{Darbyshire-2019} and \cite{newport-2019}
for more background on workplace communication.

StackOverflow is where coders go to learn, share
knowledge and build their careers. It has more than 50 million
visitors each month, and it is divided into numerous communities
defined by topics, such as ``Theoretical Computer Science''
\cite{stackoverflow-tcs}, ``Ask Ubuntu'' \cite{stackoverflow-ubuntu}
and ``\TeX -- \LaTeX'' \cite{stackoverflow-tex} to name just a few
from the 15,000 communities.

StackOverflow requires a certain etiquette, such as clarity of
questions, and reproducibility of issues described (in sufficient
detail).  Read \cite{stackoverflow-questions} on how to ask a good
question, and \cite{stackoverlfow-answers} on how to write a good
answer.  A companion to the CSUCI StackOverflow site is a LinkedIn
group on all matters AWS Cloud that is concerned with jobs, careers
and networking in the area of AWS Cloud Computing \cite{linkedin}.

\section{Building WP on AWS}

In this section we describe the backend work, where you provision an
AWS virtual server, called an EC2 instance, where EC2 abbreviates
``Elastic Cloud Compute,'' and set up a LAMP stack on it that is
needed to have a web server, and finally a WP installation on
top of that.

\begin{figure}[h]
\begin{center}
\includegraphics[width=4cm]{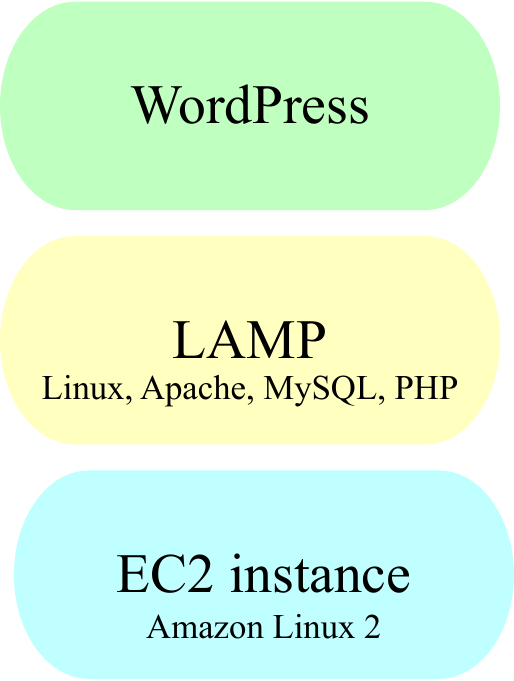}
\end{center}
\caption{WP stack}
\end{figure}

Keep in mind that there are automated ways to install WP on
AWS, as for example with AWS Lightsail
(\cite{aws-wordpress-lightsail}), or Bitnami's public AMIs with
WP (\cite{bitnami-wordpress}). However, it is important for the
purpose of this course to be able to build WP from scratch, as
this gives invaluable insights into the stack, as well as more
flexibility in configuring it. 

\subsection{EC2 with LAMP}\label{sec:lamp}

First launch an {\em Amazon Linux 2 AMI} EC2 instance; detailed
instructions are given in~\cite{aws-launch-instance}.  For the {\em
instance type}, it is recommended that you select ``General purpose''
\verb|t2.micro|, which is ``free tier eligible,'' and while it is not
a powerful server, it is enough for the purpose of the exercise. Note
that {\em instance type} can always be updated later to more powerful
specs, if required.

Once your Amazon Linux 2 EC2 instance is up and running, the next step
is to install on it a LAMP stack. ``LAMP'' here stands for Linux, the
Operating System of the instance, Apache web server, MariaDB (which is
a community-developed fork of MySQL) and PHP. Follow the instructions
here: \cite{aws-tutorial-lamp}. Note that these LAMP instructions are meant
for the specified AMI: Amazon Linux 2.

\subsubsection{Create an image}\label{sec:image}

Once the EC2 server with LAMP has been built, it should be saved as an
{\em image} before moving on to the next section
(Section~\ref{sec:wordpress}). This way, if something goes wrong
during the WP setup, the process can be restarted from this point,
rather than from scratch.  This is a transformational lesson of cloud
computing: infrastructure should be treated as code, {\em
infrastructure as code}, and the user should not be afraid to delete,
start anew, and in general approach the infrastructure as if it were a
script.

In order to launch a template, select the desired EC2 instance in the
console, select ``Actions'' and then select ``Create Image.'' This
image will be placed in ``IMAGES'' in the left margin menu, under
``AMIs.'' Now, you can launch a fully fledged EC2 server with LAMP
from that image. The ID of the AMI will be of this form:
\verb|ami-0a1b2c3d4e5f6g7h8|, and it is unique, and it can be shared
with others.

\subsection{WP}\label{sec:wordpress}

In this section you will learn how to set up WP on your EC2
instance with LAMP, as provisioned in Section~\ref{sec:lamp}.  The set
up included installation and configuration, as well as securing the
WP; security is a foundational concern --- for example, you do
not want hackers to deface your company's web page and thereby send a
message to the community that you are an easy target and thus not to
be trusted. In order to set up WP, follow these instructions:
\cite{aws-tutorial-wordpress}.

Once again an image should be created at this point as described in
Section~\ref{sec:image}, so that the user will also have an EC2 server
instance with LAMP and a fresh installation of WP.

\section{Social Media integration}

Social media is no panacea in terms of generating engagement with
ones' stakeholders. In fact, two decades into a wide-spread usage of
social media, we see that it is not all good. See for example the
following articles: \cite{kheriaty-2017,newport-2019,chalk-2019}, and
the recently published book by Hasson, {\em The Manipulators},
\cite{hasson-2020}, which shows that purveyors of social media are
prone to use it to further their own agendas.

Yet, social media can be wielded successfully to amplify the message of
a company. At the very least, it can be used to advertise a post on
various platforms; of course, the success of this strategy depends on
an earlier success in bringing a receptive audience to those platforms
in the first place.

In the following three sections we are going to show the technical
side of automating the notification about posts on three social media
platforms: Twitter, LinkedIn and Slack. In all cases this is done with
APIs, Application Programming Interfaces, which are the glue that hold
the Cloud together.

\subsection{Twitter}\label{sec:twitter}

Twitter integration automates tweeting a message about a post. That
message can be customized in WP in the final stages of
publishing a post, but in general the tweet will include a title to
the post, and a link to the post itself.  Authors often modify the
tweet by using key \#'s ({\em hashtags}) and @'s ({\em mentions}) in
order to draw more viewers.

There are various ways to accomplish twitter integration; for example
the Jetpack plugin has this integration built in (it still has to be
customized and API tokens generated). We are going to install Jetpack
in Section~\ref{sec:measuring}, but integration can also be done with
a lightweight plugin such as ``WP to Twitter.'' It is lightweight
when compared to Jetpack, as Jetpack has a lot of functionality. A
{\em plugin} is a software bundle that can be install in WP to
expand its functionality. Examples of popular plugins are ``Akismet
Anti-Spam,'' which protects WP comments from spam, and Jetpack
itself.  More information, as well are a download, of ``WP to
Twitter'' can be found here:~\cite{wordpress-twitter}.

Whether using Jetpack or ``WP to Twitter,'' there are three steps in a
typical twitter integration: First, apply to upgrade your Twitter
account to a developer account; the Twitter Developer account allows
you to access Twitter's API, which is needed to automate tweets when
posting, and it is free.  Second, download and install ``WP to
Twitter'' plugin from its page (\cite{wordpress-twitter}). Third,
configure the plugin: ``WP to Twitter'' is very well documented and
walks the user through the installation. As a valuable tip for the the
installation, make sure that the tokens are generated and entered
correctly.

\subsection{LinkedIn}\label{sec:linkedin}

LinkedIn is a professional networking site, designed to make business
connections, share resumes and find jobs. Just like Twitter and
Facebook it allows members to post, comment and like. It has two types
of posts: a short post message in the style of Twitter or Facebook,
and a longer post called an {\em article}.

Just as in the case of Twitter integration discussed in the previous
section, LinkedIn integration is achieved through plugins; the
integration is included in Jetpack, but it can also be established
with ``WP LinkedIn Auto Publish,'' which can be accessed 
here~\cite{wordpress-linkedin}.

\subsection{Slack}\label{sec:slack}

WP Slack is achieved through an RSS application in Slack. 
RSS, which stands for {\em Real Simple Syndication}, is a web feed that
allows applications to access updates to a website in a standard way.
It allows to have news aggregators in one place; for example,
\url{https://wordpress.com/read} allows one to follow various other
blogs in one place, precisely through RSS. Slack integration allows to
choose the channel where the posts will be forwarded.

\section{Measuring effectiveness}\label{sec:measuring}

In this section we are going to consider how to measure the
effectiveness of our online presence. We are going to deploy two
tools; both of them can be installed as plugins in WP, but both
require setting up an online account:
\begin{itemize}
\item {\bf Jetpack}, site: \cite{wordpress-jetpack}, and plugin: 
\cite{wordpress-jetpack-plugin}
\item {\bf Google Analytics},
site: \cite{wordpress-google}, and plugin:
\cite{wordpress-google-plugin}
\end{itemize}

Jetpack is a comprehensive plugin that includes security protections,
backup, malware scanning, and Analytics. It is the creation of a
company named {\em Automattic}, which is known for running
WP.com (a self-hosting site) and contributions to the WP
software. Jetpack is able to measure the performance of each blog post
and each page individually.

Google Analytics provides a comprehensive view of the performance of
the entire WP site. It does not have a breakdown by post and
page as Jetpack, but on the other hand it offers a lot of parameters
that Google with its vast Analytics experience can offer. For example,
{\em bounce rate}, which gives the percentage of sessions where the
viewer looks at a page, but has no other interaction with it; {\em
average session duration}, number of {\em sessions} and number of {\em
sessions per user}, where a session is a concrete period during which
a user is engaged with a web site. For example, the higher the number
of sessions per user, the more returning --- rather than new --- users
are engaged. The following YouTube video:
\cite{google-analytics-youtube} is a simple introduction to Google
Analytics.

\begin{figure}[h]
\begin{center}
\includegraphics[width=8cm]{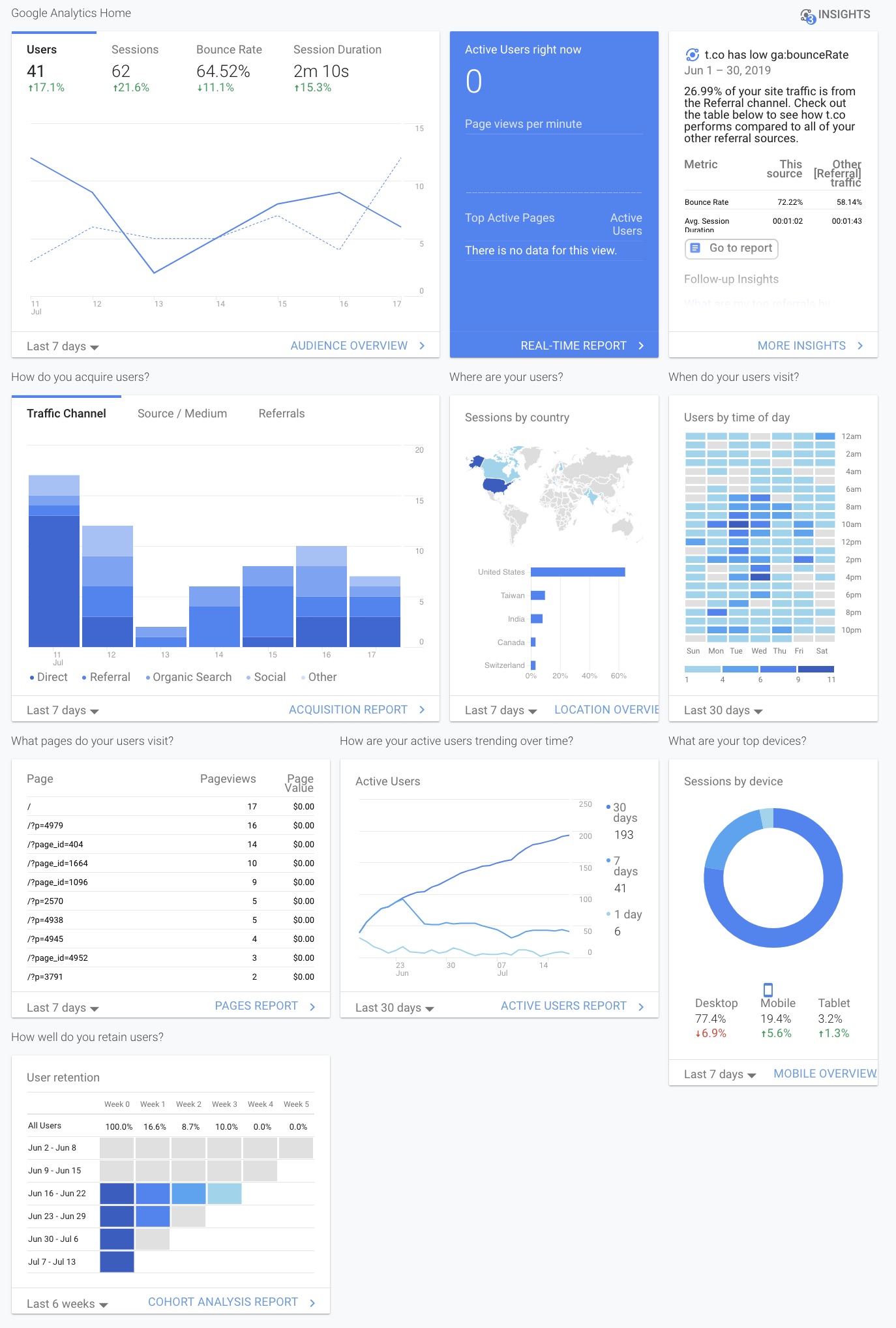}
\end{center}
\caption{Google Analytics front page for
\url{https://prof.msoltys.com}}
\end{figure}

\section{Maintenance}

\subsection{Backups}\label{sec:backup}

It is irresponsible to build a WP site without a well
understood backup procedure in place. If you are running a
professional production version of WP, you should discuss with
management two parameters related to backups: {\em Recovery Point
Objective} (PRO) and {\em Recovery Time Objective} (RTO). 

The RPO depends on how much data is your company willing to lose
(i.e., frequency of backups), and the RTO depends on how long is your
company willing to wait to recover from a catastrophic failure.

In order to meet the RPO and RTO, whatever they may be, two types of
backups are recommended. One is to do periodic snapshots of the EC2
instance and attached EBS volumes, as the site can be recovered from
such a snapshot very quickly. Still, it has to be decided how many
such snapshots to keep (what if the last ten have been affected by the
same ransomware?) and how frequently to make them; the decisions will
have an associated storage price.  

Another type of backup is to copy files from the WP
installation into into S3, for example using \verb|aws sync|
(\cite{aws-cli-sync}). Here S3 versioning might be useful, but a
decision has to be made as to how many version to keep, and at which
point --- if ever --- to move them for very cheap storage to Glacier.
The copying can be scheduled with \verb|crontab| (\cite{crontab}), by
putting the command:
\begin{verbatim}
aws s3 sync /var/www/html/ s3://<bucket-name>
\end{verbatim}
inside a shell file and invoking it at regular times from
\verb|crontab|. Note that for the EC2 instance to have access to the
S3 bucket, it is important that it assume a role with the following
permission attached: \verb|AmazonS3FullAccess|. Of course, granting
this permission violates the best practice of granting ``minimal
permission needed to do the job.'' The permission could be weakened
with a bespoke policy that grants access only to the particular S3
bucket.

\subsection{Scripts}

As the size and complexity of a system under your administration
grows, it is crucial to automate service managements with scripts. AWS
has a host of command line utilities that help with that: AWS
Serverless Application Model (AWS SAM) Local, AWS Tools for
PowerShell, AWS Amplify, as well as of course the AWS CLI (v2
recommended over v1; see \cite{aws-cli}).  To the AWS script tools we
also add the following three convenient utilities: Makefile, Bash (or
your favorite shell-script; for example, in macOS Catalina the default
shell is zsh), and a WP CLI called WP-CLI (\cite{wp-cli}).

There are certain files on your WP that change (i.e., are
updated) frequently. A resume in the ``About Me'' page would be a good
example. Instead of uploading multiple versions of this file, it is
convenient to quickly update it as it changes locally. Here is an
\href{https://gist.github.com/michaelsoltys/e54a0933efa2f2406b5fb97eab117c46}{example
of a Makefile} that can be used for that. In that example we use
\verb|scp| (secure copy) to upload the new version; note that this
will typically ask for a password. In order to automate that, the user
can export their public \verb|id_rsa.pub| from the local machine to
the EC2 instance running WP and append it to the file
\verb|.ssh/authorized_keys| there. If the local \verb|.ssh| folder
does not contain the file \verb|id_rsa.pub|, it can be easily generate
with the command \verb|ssh-keygen|. See \cite{ssh-keygen} for more
details. Also note that automating the ssh access to the EC2 instance
running your WP is a good idea in general, as it will help with
management of the instance (e.g., running \verb|sudo yum update|
frequently to
update the system). A best practice in this case is to ensure that the
Security Group that controls the ssh access to the machine is
configured to allow access from few select IPs, for example home IP
and office IP. It should definitely not be \verb|0.0.0.0/0,::0|, i.e.,
open to the world.

Here is an example of a
\href{https://gist.github.com/michaelsoltys/4c14f7f612ffc6ff3bc625501eeceb18}{Bash
script}
that is used to update a WP site, so that new plugin,
themes, and the WP core itself, can all be updated
automatically by running this script, which we call
\verb|genupdate.sh|. Note that the script also updates the
\verb|wp-cli| utility itself, keeping a copy of the older version just
in case the update breaks any current configuration.

If you are running several WP sites, or simply want to
streamline the maintenance of your WP site(s), you should
consider using \verb|wp-cli| which will speed up maintenance and
plugin/theme installation/maintenance considerably. Finally, there are
documents which are often updated and then uploaded to your WP
site. We will show how to stream line this as well.  WP-CLI is a
WP Command Line Interface (\cite{wp-cli}). 

\subsection{Security}

This is a comprehensive reference on AWS security:
\cite{soltys-aws-security}.  Note that document
\cite{aws-tutorial-lamp} we followed for building a LAMP stack in
Section~\ref{sec:lamp}, has a ``Step 3: Secure the Database Server.''
That is an important component of the setup, and if you have not
completed that step already, now would be a good time to do so. If
your WP installation has been hacked, you may find this article
helpful: \cite{wp-ec2}, where the author points out that WP'
popularity makes it a huge target, and that the login page (for
administrative maintenance) and the plugin directories are especially
voulnerable to attack.

\subsection{Cost}

Start by visiting the billing dashboard in the AWS Console, as well as
\cite{aws-billing}. It is absolutely crucial, especially at the
beginning of hosting your WP site on the AWS Cloud
infrastructure, to set up billing alarms that notify the
administrator, using AWS Simple Notification Services (SNS), that
certain cost thresholds have been reached. This will help you avoid
surprises in your monthly billing, as well as facilitate
trouble-shooting and optimizing your setup. See here for a
comprehensive list of cost tools: \cite{aws-cost-reporting}.

\subsection{Performance}

The basic setup discussed so far, depending slightly on which type you
selected in Section~\ref{sec:lamp} (\verb|t2.micro| was suggested),
can carry a website with a few simultaneous users, say~20. But if you
are serving an audience of a million users, with, say, up to 10,000
simultaneous connections to your site, you will need something more
powerful.

Enter {\em load balancers} (\cite{aws-elb}), {\em auto-scaling}
(\cite{aws-as}), CloudFront (\cite{aws-cloudfront}), {\em
fault-tolerance} and {\em Transport Layer Security} (TLS) for
encrypted (HTTPS) connections to your website. You may also want to
give your server a unique domain name, in which case you will need to
work with Route 53 (\cite{aws-route53}). See the Appendix section
``Other Deliverables'' to implement the above.

\section{Acknowledgements}

We are grateful to Rihan Pereira (CSU Channel Islands) for suggesting
the ``Blue/Green Desployment Strategy'' deliverable (in the Section
``Advanced Deliverables''), and to Kevin Williams (CSU Fullerton) for
valuable comments on the manuscript.

\appendix

\section{Deliverables}

This section contains sample assignments, called ``deliverables'',
based on the material in the paper.

\begin{enumerate}
\item  {\bf Deliverable:} Sections 1 and 2.  Log into Slack and
StackOverflow. Post a comment in Slack. Throughout the course, you
will be responsible for participating on Slack, and posting 2 {\em
good} questions on StackOverflow, and 4 {\em good} answers on
StackOverflow. Note that the StackOverflow activity should be spread
throughout the course, i.e., it should not be all done for the first
deliverable. Submit a link to your Slack comment, as well
as a PDF of your updated resume and a text file containing an ``About
Me'' page --- you will be putting both as your initial content on your
WordPress blog.

\item  {\bf Deliverable:} Section 3. Build an EC2 Amazon Linux 2
server with Wordpress installed and secured on it, by following
closely the instructions cited in the article. This will be a good
opportunity to post on StackOverflow as you encounter --- and deal
with --- technical difficulties.  Submit a text file
containing the AMI Ids of both your ``EC2 with LAMP,'' and your ``EC2
with LAMP and WordPress.'' Also submit two links: one to your ``About
Me'' page, and one to your CV (on your blog).

\item  {\bf Deliverable:} Section 4. Integrate your WordPress site
with social media, in particular with Twitter and LinkedIn. This means
that you will set up APIs (Application Processing Interface) for both
Twitter and LinkedIn (which means that you must have active accounts
with both. Create them, if you do not; you can always close them after
the course is over). Now write a post and publish it on your blog, and
it should automatically post on Twitter and LinkedIn. Submit three
links: a link to your post; a link to the corresponding tweet; and a
link to the corresponding LinkedIn post.

\item  {\bf Deliverable:} Section 5. Install Jetpack (free version)
and Google Analytics in order to measure the effectiveness of your
online campaign. Conduct an experiment to measure engagement; for
example, write a post and observe using Jetpack and Google Analytics
what kind of traction it is getting. Submit a one page write
up (as a PDF) on your experiment. Part of your experiment should be to
determine who is your audience; in Google Analytics you can see
{\em Demographics}, {\em Geolocation}, {\em Technology}, etc.; use
those datapoints to construct a narrative about your audience.

\item  {\bf Deliverable:} Section 6 and 7. Set up a well designed
backup of your site; as explained in Section~6.2, use snapshots, S3
and crontab to accomplish this. Once you are complete, you should
estimate the total monthly cost of your WordPress setup, including
everything (not just the cost of the backups) in that analysis. Set up
a billing alarm that notifies you by email when the costs have reached
50\%\ of your estimated monthly expenses, and then again when the
costs have reached 100\%\ and 150\%.  Describe your backup and cost
analysis in a two-page document.
\end{enumerate}

\section{Advanced Deliverables}

\begin{enumerate}\setcounter{enumi}{5}
\item {\bf Deliverable:}  Section 6.3. Deploy the following scripts:
\verb|wp-cli| to maintain, especially update, your WP
installation. Set up seamless ssh-ing into your EC2 instance with
\verb|ssh-keygen| and allow only select ssh traffic (port 22) to your
EC2 via its Security Group. The HTTP(S) traffic (ports 80 and 443)
should be open to the world (i.e., \verb|0.0.0.0/0|).

\item {\bf Deliverable:} Using Route53, as well as Load Balancers, set
up a domain name for your EC2 instance (note that this will require a
small payment, and may not be doable with an ``AWS Educate'' account),
and using Load Balancers set up an SSL/TLS connection to your instance
by generating a Certificate using AWS Certificate Manager.  What is
the advantage of having the load balancer encrypt and decrypt SSL/TSL
traffic, and communicating internally via HTTP with your instance? In
fact, allowing SSL/TLS connections to your EC2 instance is crucial in
order to allow secure administration of your site.

\item {\bf Deliverable:} Section 6.1. Suppose that your site is
getting a lot of traction, and you need to accommodate the higher
traffic. In order to do that, implement the following change: instead
of hosting the MySQL DB on your EC2 instance, launch an AWS RDS
(Relational Database Service), and move the DB there
(\cite{wordpress-rds}). Make sure that
you make the appropriate changes in \verb|wp-config.php| where, for
example, you will need to change the \verb|localhost| in
\verb|define('DB_HOST', 'localhost');| with the hostname of your new
RDS DB. The next deliverable is a continuation of this one.

\item {\bf Deliverable:} Suppose that a lot more media than you
expected, both audio and video, is uploaded to your WP site,
and you are running out of the provisioned capacity on your EBS
volume. Since you realized that EBS storage can get expensive, you
decide to actually shrink the EBS volume so it only holds the EC2
instance OS and the WP installation, but the
\verb|wp-content/uploads| page is going to reside in an S3 bucket; how
would you do that?  With this new setup, can you now have a load
balancer, and an auto-scaling group that spawns new EC2 instances that
are replicas of the current EC2 instance, and shuts them down as
needed?

\item {\bf Deliverable:} WP releases its software frequently
(\cite{wordpress-releases}) and you are going to implement a Blue/Green
deployment (see \cite[pg~26]{aws-cicd}) strategy to update your site
without downtime.  In order to do that, you are going to have two
target groups, the {\em Blue Group} and the {\em Green Group}, or put
another way, two environments, the {\em Blue Environment} and the {\em
Green Environment}. When a new update is released (checking for updates
could be done with \verb|crontab| and the WP CLI \verb|wp core update|), 
and when an update is available, it is deployed into the Blue
Environment, which now becomes Green, and the old Green becomes Blue,
and the Blue environment is kept idle in case a rollback is needed.
\end{enumerate}

\nocite{*}
\bibliographystyle{IEEEtran}
\bibliography{soltys_communication}

\end{document}